\documentclass[prl,aps,showpacs,twocolumn,superscriptaddress,amsmath,amssymb,floatfix]{revtex4}
\usepackage{graphicx}
\usepackage{wrapfig}
\usepackage{color}

\newcommand{\be}{\begin{equation}}
\newcommand{\ee}{\end{equation}}
\newcommand{\xx}{{\mathbf x}}

\newcommand{\kk}{{\mathbf k}}

\newcommand{\meV}{~{\rm meV}}

\begin{document}
\title{Hydrodynamic nucleation of vortices and solitons in a resonantly excited polariton superfluid}
\author{S. Pigeon}
\affiliation{Laboratoire Mat\'eriaux et Ph\'enom\`enes Quantiques,
Universit\'e Paris Diderot-Paris 7 and CNRS, UMR 7162, 75205 Paris
Cedex 13, France} 
\author{I. Carusotto}
\affiliation{INO-CNR BEC Center and Dipartimento di Fisica, Universit\`a
di Trento, I-38123 Povo, Italy}
\author{C. Ciuti}
\email{cristiano.ciuti@univ-paris-diderot.fr}
\affiliation{Laboratoire Mat\'eriaux et Ph\'enom\`enes Quantiques,
Universit\'e Paris Diderot-Paris 7 and CNRS, UMR 7162, 75205 Paris
Cedex 13, France}

\begin{abstract}
We present a theoretical study of the hydrodynamic properties of a quantum gas of exciton-polaritons in a semiconductor microcavity under a resonant laser excitation. The effect of a spatially extended defect on the superfluid flow is investigated as a function of the flow speed. The processes that are responsible for the nucleation of vortices and solitons in the wake of the defect are characterized, as well as the regimes where the superfluid flow remains unperturbed. Specific features due to the non-equilibrium nature of the polariton fluid are put in evidence.
\end{abstract}

\maketitle
In the last decades, degenerate quantum gases and liquids have been a very active field of investigation.
This research originated from the early experimental observation of superconductivity in metals and superfluidity of liquid Helium~\cite{PinesNozieres} and has experienced a further boost in the mid-90's following the experimental advances in the preparation and manipulation of quantum degenerate gases of ultra-cold bosonic and fermionic atoms~\cite{SS-LP}.

In the meanwhile, the long lasting quest for Bose-Einstein condensation of quasi-particles in solid-state systems has been finally accomplished with the observation of condensation in gases of magnons~\cite{demokritov,giamarchi}, exotic excitons in quantum Hall bilayers~\cite{eisenstein}, and exciton-polaritons in semiconductor microcavities~\cite{exc-pol-BEC,exc-pol-BEC_y}.
Motivated by recent theoretical proposals~\cite{light_superfl,light_superfl2,pol_superfl}, researchers have then started investigating the peculiar superfluidity properties of the quantum fluid of polaritons: robust propagation of a coherent polariton bullet hitting structural defects~\cite{amo_madrid} has been demonstrated, as well as a strongly enhanced lifetime of supercurrents in a polariton condensate~\cite{madrid2}. 
In the simplest case of a resonantly pumped polariton gas, the Landau criterion for frictionless flow in the presence of weak defects has been experimentally demonstrated in a quantitative way in Ref.~\cite{amo_paris}. In the same work, a Cherenkov-like conical wake of phonons was observed in the density profile when the polariton gas hits the defect at higher speed. These observations are in full agreement with the theoretical anticipations of Ref.~\cite{pol_superfl}. 

While liquid Helium experiments have given solid evidence of a critical speed for frictionless flow, they can only offer a limited access to the microscopic details of the friction process and, in particular, of the role played by vortices. Pioneering theoretical work has in fact anticipated the onset of an additional friction mechanism involving nucleation of vortex pairs and/or solitons at the surface of a spatially extended and strong defect even at speeds below the Landau critical velocity for phonon emission~\cite{frisch,josserand,jackson,kamchatnov}.

Experiments with ultra-cold atom hitting the repulsive optical potential of a blue-detuned laser have provided clear evidence for a threshold-like behavior of the friction force with a critical velocity definitely lower than the one predicted by the Landau criterion~\cite{ketterle}, and have observed the phonon wake in a supersonically moving superfluid~\cite{JILA+noi}. However, no direct evidence of hydrodynamic nucleation of vortices by a strong defect has been reported yet. The reason most probably lies in the geometry of the systems considered so far -- a very elongated condensate in Ref.~\cite{ketterle}, a radially expanding one in Ref.~\cite{JILA+noi}.

Spontaneous appearance of spatially pinned vortices in polariton condensates has been recently reported in Ref.~\cite{lagoudakis} as a result of a complex interplay of disorder, interactions, pumping and dissipation. Related work on vortices in polariton condensates has appeared in Ref.~\cite{vortex_pol_exp,vortex_pol_th}. Theoretical work on the spontaneous appearance of vortices in polariton condensates was reported in Ref.~\cite{keeling}, but the nucleation mechanism did not involve hydrodynamic instabilities at the surface of defects.

\begin{figure*}
\begin{center}
\includegraphics[width=1.6\columnwidth]{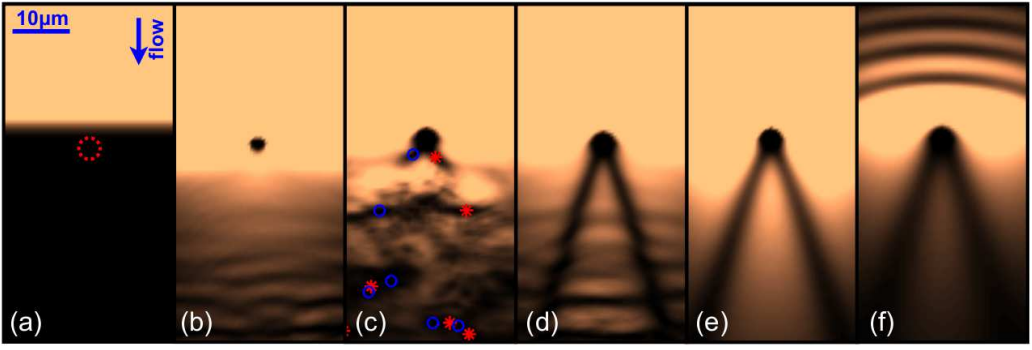}
\caption{
Panel (a): spatial profile of the pump intensity $|F_{p}(\xx)|^2$; the red dashed circle indicates the position and size of the defect. 
Panels (b-f): Normalized real-space photonic density for different values of the pump detuning $\delta_p=\hbar\omega_p-\hbar\omega_{LP}(\kk_p)$ and reservoir pump amplitude $F_{p}^{max}$ that correspond to increasing polariton densities. More specifically: $\delta_p = 0.5\meV$, $F_{p}^{max}/\gamma_C = 5.10^5~{\rm \mu m^{-1}}$ (b); $\delta_p = 0.5\meV$, $F_{p}^{max}/\gamma_C = 250~{\rm \mu m^{-1}}$ (c); $\delta_p = 0.4\meV$, $F_{p}^{max}/\gamma_C = 250~{\rm \mu m^{-1}}$ (d); $\delta_p = 0.3\meV$, $F_{p}^{max}/\gamma_C = 250~{\rm \mu m^{-1}}$ (e); $\delta_p = 0.1\meV$, $F_{p}^{max}/\gamma_C = 20~{\rm \mu m^{-1}}$ (f).
The density patterns are stationary in time in all panels except (c). Numerical simulations were performed on a $256 \times 128$ grid. The pump wave-vector has a magnitude $k_p=1\,\mu\textrm{m}^{-1}$ and is directed along the negative $y$ axis. System parameters, $\hbar\omega_{X}(k=0)= 1479\meV$, $\hbar\omega_{C}(k=0) = 1483\meV$), $\hbar\gamma_{X,C} = 0.02\meV$, and $\hbar\Omega_R=2.65 \meV$. Cavity photon mass $m_C= 40.10^{-6}{\rm m_e}$. Exciton mass $m_X$ is taken as infinite. The defect potential is a purely photonic one of depth $V_C=20\meV$. Note that the color scale is slightly saturated.
}\label{regime}
\end{center}
\end{figure*}

In the present Letter we present a theoretical investigation of the behaviour of a polariton superfluid when hitting a spatially extended defect. We specifically address the quantum hydrodynamic processes that are responsible for friction: depending on the flow speed,  the polariton gas can either flow  almost unperturbed around the defect, show a weak Cherenkov-like modulation pattern as the weak defect case~\cite{pol_superfl}, or experience the nucleation of vortices and/or solitons at the surface of the defect.
Differently from previous works on Helium and atomic gases~\cite{frisch,josserand,jackson,kamchatnov}, our calculations have to fully take into account the non-equilibrium nature of polariton gases, i.e. the need of a continuous pumping to compensate for the unavoidable polariton losses. 
Our attention will be specifically focussed on the resonant pump configuration that was adopted in Ref.~\cite{amo_paris} which offers a superior quantitative control over the experimental parameters. The case of incoherently pumped polariton condensates is presently under investigation in other groups~\cite{michiel}.

We describe the dynamics of the coherent, exciton and cavity-photon fields $\psi_{X,C}(\xx,t)$ in the two-dimensional plane of the microcavity by means of the following modified Gross-Pitaevskii equations~\cite{ciuti_review,pol_superfl}:
\begin{multline}
i \frac{d}{dt} 
\begin{pmatrix} \psi_C  \\ \psi_X 
\end{pmatrix} =  
\begin{pmatrix} F_{p}(\xx) \\ 0 
\end{pmatrix} e^{i(\textbf{k}_p\xx - \omega_p t)}
+ \\
+\begin{bmatrix}
\textbf{h}^0+\begin{pmatrix}
V_C(\xx)  & 0\\
0 &   g |\psi_X |^2
\end{pmatrix}
\end{bmatrix}
\begin{pmatrix} \psi_C\\ \psi_X 
\end{pmatrix}\label{GP}
\end{multline}

The single-particle evolution in the planar cavity is summarized by the matrix
\be
\mathbf{h^0} =  \begin{pmatrix}
\omega_C(-i\nabla) -  i\dfrac{\gamma_C}{2} & \Omega_R\\
\Omega_R & \omega_X(-i\nabla) - i \dfrac{\gamma_X}{2},
\end{pmatrix}
\ee
where $\omega_{X(C)}(\kk)$ is the dispersion of the excitons (cavity photons) as a function of in-plane momentum $\kk$ and $\gamma_{X(C)}$ is the decay rate of the excitons (cavity photons), and $\Omega_R$ the vacuum Rabi frequency of the photon-exciton coupling.  $V_{C}(\xx)$ is the photonic potential due to the defects in the sample; no potential is instead assumed to act on the exciton, $V_X=0$. The exciton-exciton interactions are described by a local interaction potential proportional of coupling constant $g$.
$F_{p}(\xx)$, $\hbar \mathbf{k}_p$ and $\hbar \omega_p$ are the spatially-dependent amplitude, momentum and energy of the pump field, respectively.   

The flow velocity of the polariton fluid is controlled by the wave vector of the pump according to $v = \hbar \mathbf{k}_p/m_{LP}$, $m_{LP}$ being the effective mass of lower-polaritons under the standard parabolic approximation of the band bottom. For a given microcavity sample and a given pump wavevector $\kk_p$, the fluid density can be determined varying either the pump detuning $\delta_p=\hbar\omega_p-\hbar\omega_{LP}(\kk_p)$ from the polariton branch or the peak pump amplitude $F^{\rm max}_p$~\cite{pol_superfl}. Within a simplest description, a value $c_s=\sqrt{\hbar g|\psi_{X}|^2/m_{LP}}$ of the speed of sound can be associated to each value of the exciton density: by tuning the value of the speed of sound with respect to the flow speed, a variety of hydrodynamic effects can be observed. The different panels of Fig.\ref{regime} correspond to different values of the detuning $\delta_p$ and of the pump amplitude $F^{\rm max}_p$.

Under a monochromatic and spatially homogenous plane-wave pump, no vortices can be observed in the polariton fluid, nor solitons. The local phase of the polariton field is in fact fixed by the pump phase, which inhibits the appearance of spatial structures such as vortices (in which the phase winds by $2\pi$ around the vortex core) or solitons (where the value of the phase has a finite jump across the density minimum). In this geometry, the effect of a large defect reduces to a simple phonon wake as discussed in Ref.~\cite{amo_paris,pol_superfl}.

A possible solution to this issue was proposed in Ref.~\cite{light_superfl2} by using a time-dependent pump: after the pump is suddenly switched off, the polariton population has a characteristic decay time $\tau\simeq \gamma_{C,X}^{-1}$ (on the order of $30~{\rm ps}$ for the system parameters chosen for the figures). During this time, the phase of the condensate is free and can develop non-trivial spatial structures as a result of the interaction with the defect.

In the present paper we shall investigate an alternative strategy based on a continuous-wave, monochromatic pump at $\omega_p$ with a non-trivial spatial profile $F_p(\xx)$. As most of the interesting dynamics of the condensate phase is taking place in the spatial region downstream of the defect, we choose a pump with an intensity profile concentrated in the half-space upstream of the defect as shown in Fig~\ref{regime}(a); in this panel, the position of the defect is marked by the red dashed circle. Polaritons are continuously injected in the cavity, propagate past the defect and extend downstream of it for a distance roughly given by $v/\gamma$: in this way, the spatial region around and past the defect shows a significant condensate density and the condensate phase is left fully free to evolve. Exception made for the polariton losses, the fluid dynamics is closely related to the standard one as given by the Gross-Pitaevskii equation~\cite{frisch,josserand,jackson,kamchatnov}.

A crucial issue of our proposal is the fine-tuning of the pump spot position: if the edge of the pumped half-space lies too far upstream of the defect, the density past the defect gets too low; if the edge overlaps too much with the defect, the phase results pinned to the pump laser one and the interesting vortex and soliton dynamics is inhibited. The intensity profile across the edge is taken with a gaussian shape. 
Different regimes of the polariton superfluid dynamics are highlighted in Fig.~\ref{regime}(b-e), from a fully superfluid regime to a Cherenkov one, passing through complex time-dependent solutions involving the periodic nucleation of vortex pairs around the equator of the defect. In the different panels we have plotted the real-space intracavity photon density profile for a given value of the pump wavevector but different values of the pump density. 
It is important to note that the polariton density on the equator of the defect is generally a bit lower than its asymptotic value in the pumped region far upstream: the appearance of solitons and/or vortices past the defect is determined by the former, while the appearance of Cherenkov precursors in the upstream region is determined by the latter. The local sound velocity in the reservoir region upstream of the defect will be denoted by $c_{sR}$, while the one on the equator will be denoted $c_{sE}$ (note that the difference between this two local sound velocity is not show in Fig.~\ref{regime}(b-e) due to the lightly saturated color scale). The defect diameter $d$ is taken to be well bigger than the healing length $\xi$, $d \approx 4\xi$. 

Existing literature on the Gross-Pitaevskii equation for atomic or Helium superfluids has predicted that a superfluid regime of unperturbed flow can survive up to $v/c_s=\sqrt{2/11}\approx 0.43$~\cite{frisch}. This regime is observed in the polariton case in Fig.\ref{regime}(b): in this case, both $v/c_{sE}$ and $v/c_{sR}$ are much smaller than one and the fluid propagation around the defect shows no trace of turbulence; the density perturbation remains very much localized in the vicinity of the defect. In this regime, the behavior of the fluid in the presence of a large defect is substantially identical to the one that is observed in the presence of a weak defect~\cite{amo_paris,pol_superfl}.

Fig.~\ref{regime}(c) corresponds to a larger value of $v/c_{sE}=0.76$. As this value is larger than the critical value $0.43$, we expect vortex nucleation from the defect surface. While each pair of vortices is dragged away in the downstream direction by the flow, new vortex pairs are continuously nucleated at the defect. In the figure, vortices with positive and negative charges are indicated by red stars and blue circles, respectively. On the other hand, no feature appears upstream of the defect, as the flow remains in this region subsonic $v/c_{sR}=0.62<1$. A video of the vortex nucleation process in this regime is available as Supplementary Material~\cite{sm}. 

For a larger value of $v/c_{sE} = 1.02$ [Fig.~\ref{regime}(e)], vortices are replaced by a pair of straight dark solitons extending past the defect at a finite angle to the flow direction. The appearance of solitons of this kind in the wake of a large defect was recently anticipated in the context of atomic condensates in Ref.~\cite{el,kamchatnov}. In the present case, the spatial inhomogeneity of the density allows for the velocity in the upstream region to remain subsonic $v/c_{sR} = 0.73$, which explains the absence of precursors upstream of the defect. 
Fig.\ref{regime}(d) shows a regime where $v/c_{sE} = 0.93$ (and $v/c_{sE} = 0.67$) and the soliton is on the edge of getting unstable towards its decay into a train of vortices. The snaky shape of the soliton is possibly a signature of a (weak) instability of this kind; analytical work for the atomic case indeed set the transition point at the slightly higher value $1$~\cite{kamchatnov}.
Increasing further the speed to a value such that $v/c_{sE} = 1.72$ and $v/c_{sR} = 1.15$, one observes an oblique soliton past the defect as well as the usual parabolic precursors propagating upstream of the defect [Fig.\ref{regime}(f)]. This result is in agreement with the predictions of~\cite{kamchatnov} for the atomic case.

\begin{figure}
\begin{center}
\includegraphics[width=0.7\columnwidth]{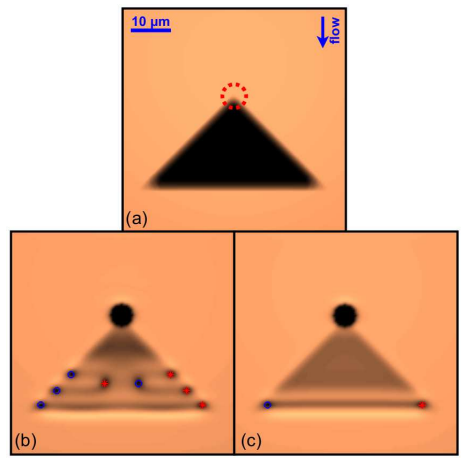}
\caption{Panel (a): real-space pump profile; the red dashed circle indicates the position and size of the defect. The external region is pumped at a value of $F_{p}^{out}$ within the vortex nucleation regime of Fig.\ref{regime}(c), $v/c_{sR} = 0.45 > 0.43$. 
The inner dark area is pumped at a lower intensity $F_{p}^{in}$. Panel (b,c): time average of the normalized photonic density for two values of the inside pump amplitude $F_{p}^{in}/F_{p}^{out}=0.1$ (b) and $0.32$ (c).  
 Calculations were performed on a $128\times128$ grid. Same system parameters as in Fig.\ref{regime}.
}\label{triangle}
\end{center}
\end{figure}

A major issue that is likely to hamper experimental observation of the vortex pairs nucleated at the surface of a large defect is the high speed at which vortices are dragged away: for typical experimental parameters, this speed is in fact on the order of $300~{\rm m.s}^{-1}$ and the size of the vortex core is $\xi \approx 0.75~{\rm \mu m}$. The time resolution that is therefore needed to neatly observe a vortex core is therefore about 10 ps.
In the following of the Letter, we shall discuss how  this problem can be overcome by a careful choice of the pump profile, e.g. the triangular one shown in Fig.\ref{triangle}(a). In the pumped external region, the phase of the superfluid is pinned to the pump phase. In the dark inner region, the phase is free to evolve and to develop vortices in the inner region. The slight jump in the polariton interaction energy at the frontiers of the pumped area contributes to the trapping of vortices along the edge. 

This effect is clearly visible in Fig.\ref{triangle}(b): vortices arrange in a regular array along the edges of the triangle. This arrangement allows for the polaritons within the triangle (i.e. in a sort of "shadow" of the defect) to have a lower flow speed as compared to the ones in the external space without breaking the irrotationality constraint of superfluid flow. Along these lines, it is immediate to understand why an increase of the fluid velocity leads to a reduced spacing between vortices. On the other hand, the total number of vortices can be increased by designing dark regions with a longer extension along the polariton flow direction.
In addition to the steady vortices located along the edges of the triangle, a few vortex-antivortex pairs are found that wander around within the triangle and, in the meanwhile, orbit around each other. Of course, no vortices can be observed in the external pumped region. Note that simulations performed with smoother edges (on the order of $\sigma = 1~{\rm \mu m}$) give a qualitatively identical physics.

\begin{figure}
\begin{center}
\includegraphics[width=0.7\columnwidth]{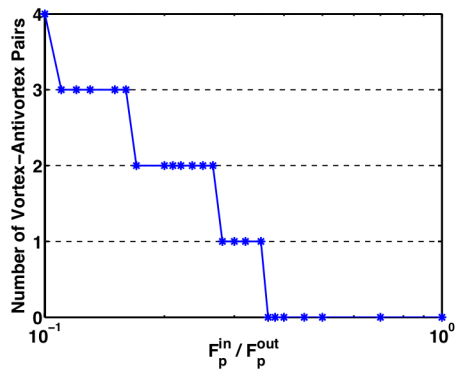}
\caption{Plot of the number of stationary vortex-antivortex pairs as a function of the inside pump amplitude. Same pump profile and system as in Fig.~\ref{triangle}.
}\label{pop}
\end{center}
\end{figure}

As a final point, it is interesting to examine what happens if the value $F_{p}^{in}$ of the pump amplitude inside the triangle is small, but finite. This physics is summarized in Figs.~\ref{triangle} and \ref{pop}; in this latter, we plot the number of observed vortices as a function of $F_{p}^{in}/F_{p}^{out}$ where $F_{p}^{out}$ is the pump amplitude outside the triangle. The limiting case $F_{p}^{in}/F_{p}^{out}=0.1\ll1$ is shown in Fig.\ref{triangle}(b) and shows both moving and trapped vortices. Increasing $F_{p}^{in}/F_{p}^{out}$ above $0.17$, a stabilization of the array is observed with the complete disappearance of moving vortices. Increasing further the inside pump amplitude, also the number of trapped pairs decreases with discrete jumps. For $F_{p}^{in}/F_{p}^{out}$ to $0.32$ a single pair of vortices at the bottom corners of the triangle is found  [Fig.\ref{triangle}(c)]. For $F_{p}^{in}/F_{p}^{out}>0.36$, the polariton phase is everywhere pinned to the incident laser one and no vortex is any longer nucleated. 

In conclusion, we have theoretically investigated vortex nucleation at the surface of a spatially extended defect in a flowing polariton superfluid. An experimentally viable protocol has been identified that allows to overcome the difficulties that stem from the non-equilibrium nature of the polariton fluid. A method to trap vortices in a regular and stationary array is illustrated. We expect that experimental studies of the vortex dynamics in polariton superfluids will shine light on fundamental aspects of the physics of superfluidity and, on a longer run, will help identifying new features that follow from the non-equilibrium nature of the polariton fluid.

We are grateful to  A. Amo, A. Bramati and E. Giacobino and M. Wouters for continuous stimulating discussions.

\end{document}